\documentclass[11pt]{article}
\usepackage{amsmath,latexsym}
\usepackage{graphicx}
\begin{document}

\title{\textbf{\textsf{Evolution of a Schwarzschild Black Hole in Phantom-like Chaplygin gas
Cosmologies}}}
\author{ \large \textbf{\textsf{Mubasher Jamil}}\footnote{Email: mjamil@camp.edu.pk}
\\ \\
%EndAName
\small
\textit{Center for Advanced Mathematics and Physics,}\\
\small\textit{National University of Sciences and Technology,}\\
\small\textit{E\&ME campus, Peshawar road, Rawalpindi - 46000, Pakistan}\\
%EndAName
}\maketitle

\begin{abstract}
In the classical relativistic regime, the accretion of phantom
energy onto a black hole reduces the mass of the black hole. In this
context, we have investigated the evolution of a Schwarzschild black
hole in the standard model of cosmology using the phantom-like
modified variable Chaplygin gas and the viscous generalized
Chaplygin gas. The corresponding expressions for accretion time
scale and evolution of mass have been derived. Our results indicate
that mass of the black hole will decrease if the accreting phantom
Chaplygin gas violates the dominant energy condition and will
increase in the opposite case. Thus our results are in agreement
with the results of Babichev \textit{et al} \cite{Babi1} who first
proposed this scenario.

\end{abstract}
\textit{Keywords}: Accretion; dark energy; Chaplygin gas; black
hole; bulk viscosity; relativistic energy conditions; phantom
energy.

\newpage
\section{Introduction}

The evidence of accelerated expansion in the observable universe is
quite compelling and has been confirmed by various astrophysical
investigations including observations of supernovae of type Ia
\cite{Riess,Perl}, anisotropies of the cosmic microwave background
radiation \cite{Page,Sper}, large scale structure and galaxy
distribution surveys \cite{Tegm}. This expansion of the universe is
supposedly driven by an exotic energy commonly called `dark energy'
possessing negative pressure $p<0$ and positive energy density
$\rho>0$, related by equation of state (EoS) $p=\omega\rho$. It
should be noted that $p=\omega\rho$ is not a true EoS for dark
energy rather a phenomenological description valid for a certain
configuration \cite{carroll}. Astrophysical data suggests that about
two third of the critical energy density is stored in the dark
energy component. The corresponding  parameter $\omega$ is then
constrained in the range $-1.38<\omega<-0.82$ \cite{Melch}. It shows
that the EoS of cosmic fluids is not exactly determined. The genesis
of this exotic energy is still unknown. The simplest and the
earliest explanation of this phenomenon was provided by the general
theory of relativity through the cosmological constant $\Lambda$.
The observational value of its energy density is $56$ to $120$
orders of magnitude smaller than that derived from the standard
theory \cite{sahni}. The satisfactory explanation of this phenomenon
requires extreme fine tuning of the cosmological parameters. Other
problem associated with $\Lambda$ is the coincidence problem (i.e.
Why did the cosmic accelerated expansion start in the presence of
intelligent beings? alternatively why the energy densities of matter
and dark energy are of the same order at current time?) which is as
yet explained either through anthropic principle \cite{wilc},
variable cosmological constant scenario \cite{jamil2} or by invoking
a dark matter-dark energy interaction \cite{campo,jamil,jamil1}. In
this context, several other models have been proposed among them are
models based on homogeneous and time dependent scalar fields termed
as quintessence \cite{Ratra}, quintom \cite{setare,ming} and
k-essence \cite{Chiba}, to name a few.

The interest in phantom energy arose when Caldwell \textit{et al}
\cite{cald} explored the cosmological consequences of the EoS,
$\omega<-1$. The dark energy can achieve this EoS if it is assumed
to be variable quantity i.e. $\omega(z)$, where $z$ is the redshift
parameter. Thus $\omega$ evolves as: for matter dominated universe
$\omega=0$, in quintessence phase $-1<\omega\leq-1/3$, for
cosmological constant dominated arena $\omega=-1$, while in phantom
regime $\omega<-1$. This scenario appears to be consistent with the
observations \cite{vikman}. In phantom cosmology, the energy density
of phantom energy will become infinite in a finite time leading the
`big rip', a kind of future singularity. Moreover, due to strong
negative pressure of phantom energy, all stable gravitationally
bound objects will be dissociated near the big rip. These findings
were later confirmed in \cite{ness} by doing numerical analysis for
the solar system and the Milky Way galaxy. In this context, the
accretion of phantom dark energy onto a black hole was first modeled
by Babichev \textit{et al} \cite{Babi1} who proved that black hole
mass will gradually decrease due to strong negative pressure of
phantom energy and tend to zero near the big rip where it will
finally disappear. Note that $\omega>-1$ leads to the opposite
scenario where black hole mass increases by accreting dark energy
until its event horizon swells up to swallow the whole universe
\cite{yuro}. Later studies \cite{noji} showed that quantum effects
dominate near the big rip singularity and consequently the mass of
black hole although decreases but stops decreasing at a finite
value. In another investigation \cite{gao}, it was demonstrated that
the physical black hole mass will increase due to accretion of
phantom energy consequently the black hole horizon and the
cosmological horizon will coincide leading the black hole
singularity to become naked, all in a finite time. This analysis has
been extended for the Riessner-N$\ddot{o}$rdstrom, Kerr-Neumann and
Schwarzschild-de Sitter black holes as well
\cite{jamil3,babi3,babi4,prado,prado1,jose1}. This result apparently
refutes the cosmic censorship conjecture (or hypothesis) which
forbids the occurrence of any naked singularity. However, the
formation of naked singularities is not completely ruled out.
Numerical simulations of gravitational collapse of spheroids show
that if the collapsing spheroid is sufficiently compact, the
singularities are hidden inside black hole while they become naked
(devoid of apparent horizon) if the spheroid is sufficiently large
\cite{shapiro1}. The future singularity of big rip is alternatively
avoided by the `big trip' where the accretion of phantom energy onto
a wormhole will increase the size of its throat so much to engulf
the whole universe \cite{moru,jamil11}. Another interesting scenario
appears in cyclic cosmology where the masses of black holes first
decrease and then increase through the phantom energy accretion and
never vanish \cite{zhang}. The implications of generalized second
law of thermodynamics to the phantom energy accretion onto a black
hole imply that accretion will be significant only near the big rip.
If this law is violated than the black hole mass will decrease
\cite{pach}. The thermodynamical investigations of phantom energy
imply its positive definite entropy which tends to become constant
if the phantom energy largely dominates the universe \cite{diaz}.
This results in the late universe to be hotter compared to the
present.

We here discuss the accretion of phantom like modified variable
Chaplygin gas and the viscous Chaplygin gas seperatly onto a black
hole. This accretion of the phantom fluid reduces the mass of the
black hole. This works serves as the generalization of the earlier
work by Babichev \textit{et al} \cite{Babi1,Babi2} who initiated the
concept of accretion of exotic matter on the black hole. We have
built our model on the same pattern by choosing more general EoS for
the dark energy.

The outline of the paper is as follows: In the next section, we
discuss the relativistic model of accretion onto a black hole. In
third section, we investigate the evolution of the mass of black
hole by the accretion of modified variable Chaplygin Gas (MVG) while
in the fourth section, we discuss the similar scenario with the
viscous generalized Chaplygin gas (VCG). Finally, we present
conclusion of our paper. The formalism adopted here is from
\cite{Babi1}.

\section{Accretion onto black hole}

We consider a Schwarzschild black hole of mass $M$ which is
gravitationally isolated and is specified by the line element (in
geometrical units $c=1=G$):
\begin{equation}
ds^{2}=\left(1-\frac{2M}{r}\right)dt^{2}-\left(1-\frac{2M}{r}\right)^{-1}dr^{2}-r^{2}(d\theta
^{2}+\sin^2 \theta d\varphi ^{2}).  \label{1}
\end{equation}
The black hole is accreting Chaplygin gas which is assumed to be a
perfect fluid specified by the stress energy tensor
\begin{equation}
T^{\mu \nu }=(\rho+p)u^{\mu }u^{\nu }-pg^{\mu \nu }. \label{2}
\end{equation}
Here $p$ and $\rho$ are the pressure and energy density of the
Chaplygin gas respectively. Due to static and spherically symmetric
nature of the black hole we assume the velocity four vector $u^{\mu
}=(u^{t}(r),u^{r}(r),0,0)$ which satisfies the normalization
condition $u^{\mu }u_{\mu }=-1$. Thus we are considering only radial
in-fall of the Chaplygin gas on the event horizon. Using the
energy-momentum conservation for $T^{\mu \nu }$, we get
\begin{equation}
ux^2(\rho+p)\sqrt{1-\frac{2}{x}+u^2}=C_1,
\end{equation}
where $x=r/M$ and $u=u^r=dr/ds$ is the radial component of the
velocity four vector $u^\mu$ and $C_1$ is a constant of integration.
The second constant of motion is obtained from $u_\mu
T^{\mu\nu}_{;\nu}=0$, which gives
\begin{equation}
ux^2\exp\left[\int\limits_{\rho_{\infty}}^{\rho_h}\frac{d\rho^\prime}{\rho^\prime+p^\prime(\rho^\prime)}\right]=-A,
\end{equation}
where $A$ is a constant of integration which is determined below for
two models of Chaplygin gas. The quantities $\rho_{\infty}$ and
$\rho_h$ are the densities of Chaplygin gas at infinity and at the
black hole horizon respectively. Further using Eqs. (3) and (4), we
obtain
\begin{equation}
(\rho+p)\sqrt{1-\frac{2}{x}+u^2}\exp\left[-\int\limits_{\rho_{\infty}}^{\rho_h}\frac{d\rho^\prime}{\rho^\prime+p^\prime(\rho^\prime)}\right]=C_2,
\end{equation}
where $C_2=-C_1/A=\rho_{\infty}+p(\rho_{\infty})$. In order to
calculate $\dot{M}$, the rate of change of mass of black hole we
integrate the Chaplygin gas flux over the entire horizon as,
$\dot{M}=\oint T^r_t dS$ where $T^r_t$ denotes the momentum density
of Chaplygin gas in the radial direction and $dS=\sqrt{-g} d\theta d
\varphi$ is the surface element of black hole horizon. Using Eqs. (2
- 5), we get this rate of change as
\begin{equation}
\frac{dM}{dt}=4\pi AM^{2}(\rho+p).  \label{3}
\end{equation}
Integration of Eq. (6) yields
\begin{equation}
M=M_{i}\left(1-\frac{t}{\tau }\right)^{-1},  \label{4}
\end{equation}
which determines the evolution of mass of black hole of initial mass
$M_{i}$ and $\tau $ is the characteristic accretion time scale given
by
\begin{equation}
\tau ^{-1}=4\pi AM_{i}(\rho+p).
\end{equation}
The number density and energy density of Chaplygin gas are related
as
\begin{equation}
\frac{n(\rho_h)}{n(\rho_{\infty})}=\exp\left[\int\limits_{\rho_{\infty}}^{\rho_h}\frac{d\rho^\prime}{\rho^\prime+p^\prime(\rho^\prime)}\right],
\end{equation}
here $n(\rho_h)$ and $n(\rho_{\infty})$ are the number densities of
the Chaplygin gas at the horizon and at infinity respectively.
Further the constant $A$ appearing in Eq. (8) is determined as
\begin{equation}
\frac{n(\rho_h)}{n(\rho_{\infty})}ux^2=-A,
\end{equation}
which is an alternative form of energy momentum conservation Eq.
(4). Moreover, the critical points of accretion (i.e. the points
where the speed of flow achieves the speed of sound
$V^2=c_s^2=\partial p/\partial \rho$) are determined as follows
\begin{equation}
u^2_\ast=\frac{1}{2x_*}, \ \ V_{*}^2=\frac{u^2_{*}}{1-3u^2_{*}},
\end{equation}
where $V^2\equiv\frac{n}{\rho+p}\frac{d(\rho+p)}{dn}-1$. Finally,
the above Eqs. (9 - 11) are combined in a single expression as
\begin{equation}
\frac{\rho_{*}+p_{*}(\rho_{*})}{n(\rho_{*})}=[1+3c_s^2(\rho_{*})]^{1/2}\frac{\rho_{\infty}+p(\rho_{\infty})}{n(\rho_{\infty})}.
\end{equation}

\section{Accretion of modified variable Chaplygin gas}

The Chaplygin gas had been proposed to explain the accelerated
expansion of the universe \cite{Kam}. It is represented by a simple
EoS, $p=-A^{\prime}/\rho$ where $A^{\prime}$ is positive constant.
The corresponding expression for the evolution of energy density is
\begin{equation}
\rho=\sqrt{A^{\prime}+\frac{B}{a^6}},
\end{equation}
where $B$ is a constant of integration. From Eq. (13) we find the
following asymptotic behavior for the density \cite{cope}:
\begin{equation}
\rho\sim\sqrt{B}a^{-3}, \ \ \ \ a\ll(B/A^{\prime})^{1/6},
\end{equation}
\begin{equation}
\rho\sim\-p\sim\sqrt{A^{\prime}}, \ \ \ \ a\gg(B/A^{\prime})^{1/6}.
\end{equation}
Thus for small $a$, it gives matter dominated era at earlier times
while for large $a$ we get dark energy dominated era at late times.
Thus Chaplygin gas has the property of giving a unified picture of
the evolution of the universe. The observational evidence in favor
of cosmological models based on Chaplygin gas is quite encouraging
\cite{Dev,silva,berto,barreiro}. The Chaplygin gas model favors a
spatially flat universe which agrees with the observational data of
Sloan Digital Sky Survey (SDSS) and Supernova Legacy Survey (SNLS)
with 95.4 \% confidence level \cite{Xun}. Consequently, various
generalizations of Chaplygin gas have been proposed in the
literature to incorporate any other dark component in the universe
(see e.g. \cite{Benaoum,Yi,set1,set2} and references therein).

We here consider an equation of state which combines various EoS of
Chaplygin gas given by \cite{debnath}
\begin{equation}
p=A^{\prime }\rho-\frac{B(a)}{\rho^{\alpha }}, \ \ B(a)=B_o a^{-m}.
\end{equation}
Here $A^{\prime }$, $B_o$ and $m$ are constant parameters with
$0\leq \alpha \leq 1$. For $A^{\prime}=0,$ Eq. (16) gives
generalized Chaplygin gas. Further if $B=B_o$ and $\alpha=1$, it
yields the usual Chaplygin gas. Also Eq. (16) reduces to modified
Chaplygin gas if only $B=B_o$. Moreover, if only $A^{\prime }=0$,
the same equation represents variable Chaplygin gas.

We consider the background spacetime to be spatially flat ($k=0$),
homogeneous and isotropic represented by Friedmann-Robertson-Walker
(FRW) metric. The spacetime is assumed to contain only one component
fluid i.e. the phantom energy represented by the Chaplygin gas EoS.
The corresponding field equations are
\begin{equation}
H^{2}\equiv\left(\frac{\dot{a}}{a}\right)^{2}=\kappa^2\rho.
\label{5}
\end{equation}
\begin{equation}
\dot{H}+H^2=\frac{\ddot{a}}{a}=-\frac{\kappa^2}{2}(\rho+3p),
\end{equation}
where $\kappa^2=8\pi/3$. The conservation of energy is
\begin{equation}
{\dot{\rho}}+3H(\rho+p)=0.
\end{equation}
Using Eqs. (16) and (19), the density evolution is given by
\begin{equation}
\rho=\left[\frac{3B_o(1+\alpha )}{\{3(1+\alpha )(1+A^{\prime })-m\}
}\frac{1}{a^{m}}+\frac{\Psi}{a^{3(1+\alpha )(1+A^{\prime
})}}\right]^{\frac{1}{ 1+\alpha }}.  \label{6}
\end{equation}
Here $\Psi$ is a constant of integration. Note that to obtain the
increasing energy density of phantom energy with respect to scale
factor $a(t)$, we require the coefficients of $a(t)$ in Eq. (20) to
be positive i.e. $\Psi\geq0$, $B_o(1+\alpha )>0$ and $3(1+\alpha
)(1+A^{\prime })-m>0$. Moreover, the exponents of $a(t)$ must be
negative i.e. $m<0$ and $3(1+\alpha)(1+A^\prime)<0$ to obtain
increasing $\rho$. These constraints together imply that
$m>3(1+\alpha)(1+A^\prime)$. Another way of getting positive $\rho$
is by setting $m>0$, $1+A^\prime>0$ and $m<3(1+\alpha)(1+A^\prime)$.
Further, using Eq. (9) the ratio of the number density of Chaplygin
gas near horizon and at infinity is calculated to be
\begin{equation}
\frac{n(\rho _h)}{n(\rho _{\infty })}=\left[\frac{\rho _h^{1+\alpha
}(1+A^{\prime })-B(a)}{\rho _{\infty }^{1+\alpha }(1+A^{\prime })-B(a)}\right]^{%
\frac{1}{(1+\alpha )(1+A^{\prime })}}\equiv \Delta_1.  \label{7}
\end{equation}
Notice that the function $B(a)$ can be expressed in terms of $\rho$
implicitly and is determined from Eq. (20). Making use of Eq. (11),
the critical points of accretion are given by
\begin{equation}
u_{\ast }^{2}=\frac{\Delta_2}{1+3\Delta_2}, \ \ x_{\ast
}=\frac{1+3\Delta_2}{2\Delta_2},
\end{equation}
where
\begin{equation}
V_\ast^2= A^{\prime }+\frac{\alpha B(a)}{\rho _{*}^{\alpha
+1}}\equiv\Delta_2
\end{equation}
Thus for the critical points to be finite and positive, we require
either $\Delta_2>0$ or $\Delta_2<0$ and $\Delta_2<-1/3$. For the
accretion to be critical, the quantity $V^2$ must become supersonic
from the initial subsonic somewhere near the black hole horizon. For
the MVG, we have $\omega=A^{\prime}-B/\rho^{1+\alpha}<0$, since
$A^\prime<-1$. One can observe that fluids having EoS $\omega<0$ are
hydrodynamically unstable i.e. the speed of sound in that medium can
not be defined since $c_s^2<0$. In order to overcome this problem
Babichev \textit{et al} \cite{babi5} proposed a redefinition of
$\omega$ with the help of a generalized linear EoS given by
$p=\beta(\rho-\rho_o)$, where $\beta$ and $\rho_o$ are constant
parameters. Here $\beta>0$ refers to a hydrodynamically stable while
$\beta<0$ corresponds to hydrodynamically unstable fluid. We will
not be interested in the later case here. Note that now two
parameters $\omega$ and $\beta$ are related by
$\omega=\beta(\rho-\rho_o)/\rho$. Further $\omega<0$ now corresponds
to $\beta>0$ and $\rho_o>\rho$ thereby making the previously
unstable fluid, now stable. We also have $c_s^2\equiv\partial
p/\partial\rho=\beta$. Since for stability, we require $\beta>0$ and
$0<c_s^2<1$, it leads to
$0<\frac{1}{\rho-\rho_o}(A^\prime\rho-B/\rho^{\alpha})<1$ and
$0<\beta<1$. Hence the EoS parameter is now well-defined with
$A^\prime<-1$ and $\rho_o>\rho$. Thus the stability of the phantom
like MVG is guaranteed with the use of generalized linear EoS.

The constant $A$ is determined from Eq. (10) to give
\begin{equation}
-A=\frac{\Delta_1}{4}\left(\frac{1+3\Delta_2}{\Delta_2}\right)^{3/2}.
\label{9}
\end{equation}
Using Eq. (8) the characteristic evolution time scale becomes
\begin{equation}
\tau^{-1}=\pi
M_i(\rho+p)\frac{\Delta_1}{4}\left(\frac{1+3\Delta_2}{\Delta_2}\right)^{3/2}.
\end{equation}
Using Eq. (25) in (7), the black hole mass is given by
\begin{equation}
M(t)=M_i\left[1-\pi
M_it(\rho+p)\frac{\Delta_1}{4}\left(\frac{1+3\Delta_2}{\Delta_2}\right)^{3/2}\right]^{-1},
\end{equation}
which determines the evolution of mass of black hole accreting
phantom MVG. It can be seen that if the phantom MVG violates the
dominant energy condition $\rho+p>0$ than mass $M$ of the black hole
will decrease. Contrary if this condition is satisfied than $M$ will
increase. Thus in the classical relativistic regime, this result is
in conformity with the result of Babichev \textit{et al}
\cite{Babi1}. We also stress here that although our metric (1) is
static, we get a dynamical mass $M(t)$ in Eq. (26). Astrophysically
the mass of a black hole is a dynamical quantity: the mass will
increase if the black hole accretes classical matter (which
satisfies $\rho+p>0$) however it will decrease for the exotic
phantom energy accretion. The mass can also decrease if the Hawking
evaporation process is invoked. Hence the static black holes may not
necessarily correspond to the astrophysical black holes. We also
stress that $\omega>0$ ($\omega<0$) corresponds to non-phantom
(phantom) MVG fluid; although the accretion through the critical
point is possible in both the cases, only phantom MVG violating the
dominant energy condition will reduce the mass of black hole.

\section{Accretion of viscous generalized Chaplygin gas}

In viscous cosmology, the presence of viscosity corresponds to space
isotropy and hence is important in the background of FRW spacetime
\cite{brevik,hu,coles}. The presence of viscous fluid can explain
the observed high entropy per baryon ratio in the universe
\cite{mis}. It can cause exponential expansion of the universe and
can rule out the initial singularity which mares the standard big
bang picture. The matter power spectrum in bulk viscous cosmology is
also well behaved as there are no instabilities or oscillations on
small perturbation scale \cite{coli}. Any cosmic fluid having
non-zero bulk viscosities has the EoS $p_{eff}=p+\Pi$, where $p$ is
the usual isotropic pressure and $\Pi$ is the bulk viscous stress
given by $\Pi\equiv\xi(\rho)u^\mu_{;\mu}$ \cite{ecka}. The scaling
of viscosity coefficient is $\xi=\xi_o\rho^n$ where $n$ is a
constant parameter and $\xi(t_o)=\xi_o$. Note that for $0\leq n\leq
1/2$, we have de Sitter solution and for $n>1/2$ we get deflationary
solutions. The viscosity coefficient is generally taken to be
positive for positive entropy production in conformity with the
second law of thermodynamics \cite{cata}. Moreover, the entropy
corresponding to viscous cosmology is always positive and increasing
which is consistent with the thermodynamic arrow of time. Infact the
cosmological model with viscosity is consistent with the
observational SN Ia data at lower redshifts while it mimics the
$\Lambda$CDM model in the later cosmic evolution \cite{hu1}. It is
proved in \cite{barr1,barr2} that FRW spacetime filled with perfect
fluid and the bulk viscous stresses will violate the dominant energy
condition.

Thus the effective pressure is given by
\begin{equation}
p_{eff}\equiv p+\Pi,
\end{equation}
where $\Pi=-3H\xi$ and $p=\chi/\rho^{\alpha }$ with $\chi $ is a
constant. Thus in the VCG case, the standard FRW equation becomes
\cite{Zhai}
\begin{equation}
\frac{\ddot{a}}{a}=-\frac{\kappa^2}{2}(\rho+3p_{eff}).
\end{equation}
Further the energy conservation principle gives
\begin{equation}
\dot{\rho}+3H(\rho+p_{eff})=0,
\end{equation}
which shows that the viscosity term serves as the source term. Using
Eqs. (17) and (27) in (29), we get
\begin{equation}
\frac{a}{3}\frac{d\rho}{da}+\rho+\frac{\chi}{\rho^\alpha}-3\kappa\xi(\rho)\sqrt{\rho}=0.
\end{equation}
Thus solving Eq. (30) we get
\begin{equation}
a(t)=a_o\exp\left[-\int\limits_{\rho_o}^{\rho}\frac{\rho^{\prime\alpha}
d\rho^{\prime}}{\rho^{\prime\alpha+1}-3\kappa\xi
(\rho^{\prime})\rho^{\prime\alpha+\frac{1}{2}}+\chi}\right]^{\frac{1}{3}}.
\end{equation}
For our further analysis we shall assume $\xi$ to be constant.

The ratio of the number density of VCG near black hole horizon and
at infinity is given by
\begin{equation}
\frac{n(\rho _{h})}{n(\rho _{\infty })}=\exp \left[\int\limits_{\rho
_{\infty} }^{\rho _{h}}\frac{\rho^{\prime \alpha }d\rho ^{\prime
}}{\rho^{\prime \alpha
+1}-3\kappa\xi\rho^{\prime\alpha+\frac{1}{2}}+\chi }\right]\equiv
\Delta_3. \label{11}
\end{equation}
The corresponding critical points of accretion are
\begin{equation}
u_{\ast }^{2}=\frac{\Delta_4}{3\Delta_4-1},\ \ x_{\ast
}=\frac{3\Delta_4-1}{2\Delta_4},
\end{equation}
where
\begin{equation}
V_{*}^2= -\left(\frac{\alpha \chi }{\rho _{\ast}^{\alpha
+1}}+\frac{3}{2\sqrt{\rho_{\ast}}}\kappa\xi\right)\equiv\Delta_4.
\end{equation}
Notice that for the critical points to be finite and positive valued
we require either $\Delta_4<0$ or $\Delta_4>1/3$. Using Eq. (11) we
see that the speed of flow at the critical point is $V^2=-\Delta_4$.
Further, the EoS parameter is $\omega=\chi/\rho^{1+\alpha}-3\xi
H/\rho$ ($=\chi/\rho^{1+\alpha}-\sqrt{3}\kappa\xi/\sqrt{\rho}$).
Note that if $\chi<0$ then $\omega<0$ and stability of VCG is lost.
However, if we here invoke the argument presented in the last
section, we can consider accretion with $\omega<0$. Using the
generalized linear EoS $p=\beta(\rho-\rho_o)$ for the phantom
energy, we obtain $\beta>0$ and $\rho_o>\rho$ for $\omega<0$. Using
the definition $c_s^2\equiv\partial p/\partial\rho=\beta$ and
stability requirements $\beta>0$ and $0<c_s^2<1$ lead to
$0<\frac{1}{\rho-\rho_o}(\chi/\rho^{\alpha}-\sqrt{3\rho}\kappa\xi)<1$
and $0<\beta<1$. The EoS parameter $\beta$ is now well-defined with
$\chi<0$ and $\rho_o>\rho$. Therefore the stability of the phantom
like VCG is assured with the use of generalized linear EoS.

Using Eq. (10) the constant $A$ is now determined to be
\begin{equation}
-A=\Delta_3\left(\frac{3\Delta_4-1}{2\Delta_4}\right)^{3/2}.
\label{13}
\end{equation}
The characteristic evolution time scale is
\begin{equation}
\tau^{-1}=4\pi
M_i(\rho+p)\Delta_3\left(\frac{3\Delta_4-1}{2\Delta_4}\right)^{3/2}.
\end{equation}
Using Eqs. (35) and (36) in (7), we get black hole mass evolution as
\begin{equation}
M(t)=M_i\left[1-4\pi
M_it(\rho+p)\Delta_3\left(\frac{3\Delta_4-1}{2\Delta_4}\right)^{3/2}\right]^{-1}.
\end{equation}
It can be seen that black hole mass will decrease when $\rho+p<0$
and increase in the opposite case. It is emphasized that this result
is valid till the contribution of viscous stress is negligible
compared to isotropic stress. For the sake of clarity, we emphasis
that fluid violating the standard energy conditions is termed
`exotic' and hydrodynamically unstable i.e. its existence is not
fully guaranteed. But this conclusion is drawn due to the `bad'
choice of the EoS ($p=\omega\rho$) in the analysis. The result is
reversed and remedied when we introduce the generalized linear EoS
in our model which makes the accretion of exotic fluid much more
practical.

\section{Conclusion}
We have investigated the accretion of two different forms of
phantom-like Chaplygin gas onto a Schwarzschild black hole. The time
scale of accretion and the evolution of mass of black hole are
derived in the context of two widely studied Chaplygin gas models
namely the modified variable Chaplygin gas and the viscous
generalized Chaplygin gas. Although the phantom energy is an
unstable fluid as it corresponds to a medium with indeterminate
speed of sound and super-luminal speeds. These pathologies arise due
to bad choices of the equations of state for the phantom energy and
hence can be removed by choosing some suitable transformation from
one EoS to another or a totally new EoS for this purpose. This work
serves as the generalization of the earlier work by Babichev
\textit{et al} \cite{Babi1}. It should be noted that we have ignored
matter component in the accretion model. Thus it will be more
insightful to incorporate the contributions of matter along with the
Chaplygin gas during accretion onto black hole. Moreover our
analysis can be extended to the case of rotating black holes as
well.

\subsection*{Acknowledgments} I would like to thank Muneer Ahmad Rashid
for enlightening discussions during this work. Useful criticism on
this work from the anonymous referee is also gratefully
acknowledged.


\small
\begin{thebibliography}{99}
\bibitem{Riess} A.G. Riess et al, 1998 \textit{Ast. J.} 116, 1009
\bibitem{Perl} S. Perlmutter et al, 1999 \textit{Ap J.} 517, 565
\bibitem{Page} L. Page  et al, 2003 \textit{Ap. J. Supp.} 148, 233
\bibitem{Sper} D.N. Spergel et al, 2003 \textit{Ap. J. Supp.} 148, 175
\bibitem{Tegm} M. Tegmark et al, 2004 \textit{Phys. Rev. D} 69, 103501
\bibitem{carroll} S.M. Carroll et al, 2003 \textit{Phys. Rev. D} 68, 023509
\bibitem{Melch} A. Melchiorri et al, 2003 \textit{Phys. Rev. D} 68, 043509
\bibitem{sahni} V. Sahni, arXiv: 0403324 [astro-ph]
\bibitem{wilc} F. Wilczek, arXiv: 0408167 [hep-ph]
\bibitem{jamil2} F. Rahaman et al, arXiv: 0809.4314 [gr-qc]
\bibitem{campo} S. Campo et al, 2006 \textit{Phys. Rev. D} 74, 023501
\bibitem{jamil} M. Jamil and M.A. Rashid, 2008 \textit{Eur. Phys. J. C} 56,
429
\bibitem{jamil1} M. Jamil and M.A. Rashid, \textit{Eur. Phys. J. C} 58 (2008) 111.
\bibitem{Ratra} B. Ratra and J.P.E. Peebles, 2003 \textit{Rev.
Mod. Phys.} 75, 559
\bibitem{setare} M.R. Setare, 2006 \textit{Phys. Lett. B} 641, 130
\bibitem{ming} M. Li et al, 2005 \textit{J. Cosmo. Astropart. Phys.}
12, 002
\bibitem{Chiba} T. Chiba et al, 2000 \textit{Phys. Rev. D} 62, 023511
\bibitem{cald} R.R. Caldwell et al, 2003 \textit{Phys. Rev. Lett.} 91, 071301
\bibitem{vikman} A. Vikman, arXiv: 0407107 [astro-ph]
\bibitem{ness} S. Nesseris and L. Perivolaropoulos, arXiv:
0410309v2 [astro-ph]
\bibitem{Babi1} E. Babichev et al, 2004 \textit{Phys. Rev. Lett.} 93, 021102
\bibitem{yuro} A.V. Yurov et al, 2006 \textit{Nuc. Phys. B} 759, 320
\bibitem{noji} S. Nojiri and S. Odintsov, arXiv: 0408170 [hep-th]
\bibitem{gao} C. Gao et al, 2008 \textit{Phys. Rev. D }78, 024008
\bibitem{jamil3} M. Jamil et al, \textit{Eur. Phys. J.} C 58 (2008) 325.
\bibitem{babi3} E. Babichev et al, arXiv: 0806.0916 [gr-qc]
\bibitem{babi4} E. Babichev et al, arXiv: 0807.0449 [gr-qc]
\bibitem{prado} P. Martin-Moruno et al, arXiv: 0803.2005 [gr-qc]
\bibitem{prado1} P. Martin-Moruno et al, 2006 \textit{Phys. Lett. B} 640, 117
\bibitem{jose1} J.A.J. Madrid and P.F. Gonzalez-Diaz, arXiv: 0510051
[astro-ph]
\bibitem{shapiro1} S.L. Shapiro and S.A. Teukolsky, 1991 \textit{Phys. Rev.
Lett.} 66, 994
\bibitem{moru} P.M. Moruno, 2008 \textit{Phys. Lett. B} 659, 40
\bibitem{jamil11} M. Jamil, \textit{Nuovo Cimento B} 123 (2008) 599.
\bibitem{zhang} X. Zhang, arXiv: 0708.1408v1 [gr-qc]
\bibitem{pach} J.A. Pacheco and J.E. Howarth,  arXiv:
0709.1240 [gr-qc]
\bibitem{diaz} P.F. Gonzalez-Diaz and C.L. Siguenza, 2004 \textit{Phys. Lett. B} 589, 78
\bibitem{Babi2} E. Babichev et al, arXiv: 0507119v1 [gr-qc]
\bibitem{Kam} A. Kamenshchik et al, 2001 \textit{Phys. Lett. B} 511, 265
\bibitem{cope} E.J. Copeland et al, 2006 \textit{Int. J. Mod. Phys. D}
15, 1753
\bibitem{Dev} A. Dev et al, 2003 \textit{Phys. Rev. D }67, 023515.
\bibitem{silva} P.T. Silva and O. Bertolami, 2003 \textit{Ap. J.} 599, 829
\bibitem{berto} O. Bertolami et al, 2004 \textit{Mon. Not. Roy. Ast. Soc.}
353, 329
\bibitem{barreiro} T. Barreiro et al, 2008 \textit{Phys. Rev. D} 78,
043530
\bibitem{Xun} W.P. Xen and Y.H. Wei, 2007 \textit{Chin. Phys. Lett. }24, 843
\bibitem{Benaoum} H.B. Benaoum, arXiv: 0205140v1 [hep-th]
\bibitem{Yi} Y.X. Yi et al, 2007\textit{ Chin. Phys. Lett.} 24, 302
\bibitem{set1} M.R. Setare, 2007 \textit{Phys. Lett. B }648, 329
\bibitem{set2} M.R. Setare, 2007 \textit{Phys. Lett. B} 654, 1
\bibitem{debnath} U. Debnath, arXiv: 0710.1708v1 [gr-qc]
\bibitem{babi5} E. Babichev et al, 2005 \textit{Class. Quantum
Gravit.} 22, 143
\bibitem{brevik} I. Brevik and O. Gorbunova, arXiv: 050401v2 [gr-qc]
\bibitem{hu} M.G. Hu and X.H. Meng, 2006 \textit{Phys. Lett. B} 635, 186
\bibitem{coles} P. Coles and F. Lucchin,\textit{ Cosmology: The origin and
evolution of cosmic structure }(John Wiley, 2003)
\bibitem{mis} C.W. Misner, 1968 \textit{Ap. J.} 151, 431
\bibitem{coli} R. Colistete et al, 2007 \textit{Phys. Rev. D} 76, 103516
\bibitem{ecka} C. Eckart, 1940 \textit{Phys. Rev.} 58, 919
\bibitem{cata} M. Cataldo et al, arXiv: 0506153 [hep-th]
\bibitem{hu1} M. Hu and X. Meng, 2006 \textit{Phys. Lett. B} 635, 186
\bibitem{barr1} J.D. Barrow, 1987 \textit{Phys. Lett. B }180, 335
\bibitem{barr2} J.D. Barrow, 1988 \textit{Phys. Lett. B} 310, 743
\bibitem{Zhai} X.H. Zhai et al, arXiv: 0511814v2 [astro-ph]

\end{thebibliography}
\end{document}